\documentclass[twocolumn,english,amssymb,aps,prl,superscriptaddress,tightenlines,notitlepage,longbibliography]{revtex4-2}
\usepackage{lmodern}
\usepackage[normalem]{ulem}
\usepackage[T1]{fontenc}
\usepackage[latin9]{inputenc}
\usepackage{color}
\usepackage{babel}
\usepackage{verbatim}
\usepackage{amsmath}
\usepackage{amssymb}
\usepackage{cancel}
\usepackage{graphicx}
\usepackage{array}
\usepackage{esint}
\usepackage{booktabs}
\usepackage[unicode=true,pdfusetitle,
 bookmarks=true,bookmarksnumbered=true,bookmarksopen=false,
 breaklinks=false,pdfborder={0 0 1},backref=false,colorlinks=true]
 {hyperref}
\hypersetup{
 linkcolor=magenta,urlcolor=blue,citecolor=blue,pdfstartview={FitH},hyperfootnotes=false,bookmarks=False}

\makeatletter
\usepackage{babel}
\date{\today}
\usepackage{xcolor}
\usepackage{graphicx}
\allowdisplaybreaks
\def\bigl{\mathopen\big}

\def\bigr{\mathclose\big}

\renewcommand{\big}{\bBigg@\@ne}
\renewcommand{\Big}{\bBigg@{1.5}}
\renewcommand{\bigg}{\bBigg@\tw@}
\renewcommand{\Bigg}{\bBigg@{2.5}}

\newcommand{\biggg}{\bBigg@\thr@@}
\newcommand{\Biggg}{\bBigg@{3.5}}

\newcommand{\before}[1]{\iffalse {#1}\fi}

\makeatother

\begin{document}
\title{Reentrant superconductivity enabled by spin-orbit coupling: Application to UTe$_2$}
\author{Changhee Lee}
\affiliation{Department of Physics and MacDiarmid Institute for Advanced Materials and Nanotechnology, University of Otago, P.O. Box 56, Dunedin 9054, New Zealand}
\author{Nico A. Hackner}
\affiliation{Department of Physics and MacDiarmid Institute for Advanced Materials and Nanotechnology, University of Otago, P.O. Box 56, Dunedin 9054, New Zealand}
\author{Daniel F. Agterberg}
\affiliation{Department of Physics, University of Wisconsin-Milwaukee, Milwaukee, Wisconsin 53201, USA}
\author{P. M. R. Brydon}
\affiliation{Department of Physics and MacDiarmid Institute for Advanced Materials and Nanotechnology, University of Otago, P.O. Box 56, Dunedin 9054, New Zealand}

\begin{abstract}
Reentrant superconductivity has been understood primarily in terms of the Jaccarino-Peter field-compensation effect or from a change of the strength in the pairing interaction. However, neither mechanism appears able to entirely explain the remarkable phase diagram of UTe$_2$. 
Here we propose a generic theory of the field-enhancement of \emph{opposite}-spin Cooper pairings which does not necessitate the coexistence of magnetism or the vicinity of a magnetic quantum critical point. 
Our analytical treatment shows that the reentrance has its origin in the interplay of the sublattice degrees of freedom and spin-orbit coupling, which can can strikingly enhance opposite-spin Cooper pairings at strong Zeeman fields. Based on these results, we show that a pairing state with $B_{3u}$ symmetry can reproduce the highly anisotropic phase diagram of the reentrant superconducting state of UTe$_2$.
\end{abstract}
\maketitle

\section{Introduction}

The suppression of superconductivity by magnetic fields is well established, with the Lorentz force and the Zeeman effect providing robust and universal depairing mechanisms. 
In a small number of superconductors, however, magnetic fields enhance the critical temperature~\cite{Aoki2019ReviewFMSC,Llanos2026}, or even induce reentrant superconductivity \cite{Uji2001,Konoike2004, Levy2005,Lewin2023,Yang2026}. Such field-enhanced superconductivity is typically explained by one of two mechanisms. The first is the Jaccarino-Peter effect, in which an applied magnetic field compensates the internal magnetic field, resulting in a reduced net field for the conduction electrons \cite{Jaccarino1962,Uji2001,Konoike2004}. The second explanation is that the applied magnetic field tunes the magnetic fluctuations responsible for the pairing interaction \cite{Aoki2019ReviewFMSC,Llanos2026,Wu2025,Hattori2013}. 

A spectacular example of this physics occurs in UTe$_2$. This material displays several superconducting phases under magnetic fields at ambient pressure~\citep{Ran2019a,Lewin2023}: a low-field superconducting state (SC1) which survives field strengths far exceeding the Pauli limit; a reentrant superconducting state (SC2) at magnetic field $15\,{\rm T}< H < 35\,{\rm T}$ along the $b$-axis; and another reentrant superconducting state (SC3) appearing in the magnetic field-polarized state. There is growing evidence that field-enhanced magnetic fluctuations play a key role in stabilizing the SC3 state \cite{Zambra2026,Wu2025}. However, such enhanced fluctuations have not been associated with the SC2 phase; Indeed, under increasing pressure, SC2 has been observed to move to lower fields and become stable at zero field~\cite{Vasina2025}. This phenomenology is not explicable by the usual mechanisms of field-enhanced superconductivity, suggesting another effect is at play. 

\begin{figure}
\includegraphics[width=0.9\linewidth]{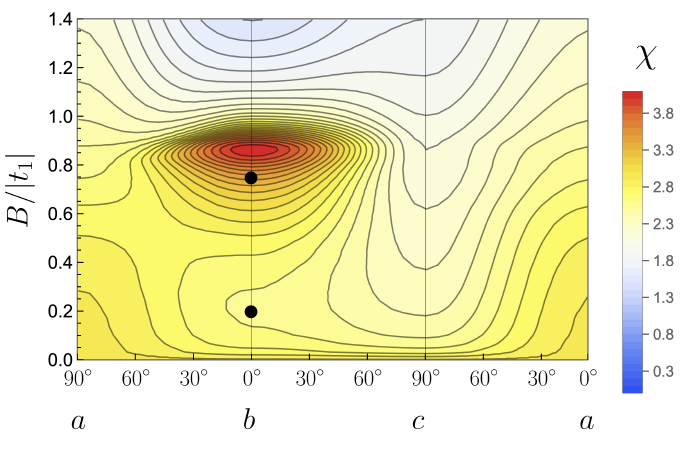}
\caption{\label{fig:Fig1}Magnetic Field-Angle superconducting phase diagram at a constant temperature $T=0.005|t_1|$ obtained by using the same model as that used for Fig.~\ref{fig:fig2}. The color gradient represents the pairing susceptibility $\chi$ of an opposite-spin spin-triplet pairing channel represented by $\tau_y s_y$. The black dot indicates the magnetic field used to obtain Figure.~\ref{fig:fig3}(b)}\end{figure}

Motivated by the SC2 phase of UTe$_2$, we propose a theory of reentrant superconductivity that does not rely on field-enhanced magnetic fluctuations or the Jaccarino-Peter effect. At the heart of our theory is a local, spin-triplet state in which quasiparticles on the two U atoms in the unit cell form a sublattice-singlet Cooper pair.  Although the pairing interaction for this state originates from local inter-sublattice ferromagnetic fluctuations, it is stabilized as a weak-coupling instability by the spin-sublattice texture of the band states at the Fermi energy~\cite{Tatsuya2021}. We find that the interplay of the magnetic field with spin-orbit coupling can dramatically enhance the stability of this state by modifying this texture. Under rather general conditions,  this leads to a pairing susceptibility that displays a pronounced peak for fields along the $b$-axis, as shown in Fig.~\ref{fig:Fig1}. We argue that the field strength at which this peak occurs is consistent with the maximum critical temperature of the SC2 phase. We note that the same pairing state has recently been suggested as an explanation for the extremely strong fluctuations observed in the high-field superconducting state~\cite{Kamat2026}. Significantly, this high-pressure state evolves into the reentrant SC2 phase as the pressure is reduced~\cite{Kamat2026,Vasina2025}.
We conclude this work by discussing the potential relevance of our theory to other superconductors which display reentrant superconductivity, specifically the uranium-based ferromagnetic superconductors~\citep{Aoki2019ReviewFMSC}.

\section{Microscopic model}
The heavy-fermion nature of UTe$_2$ implies that  quasiparticles originating from the U~5$f$ orbitals must play a central role in the physics. A minimal model for the low-energy states should therefore explicitly include the sublattice degree of freedom arising from the dimer (or ``rung'') of $U$ atoms in the unit cell.  The general form of such a Hamiltonian is
\begin{align}
h_{\vec{k}}=\varepsilon_{0,\vec{k}} \tau_0 s_0 +(\vec{t}_{\vec{k}}\cdot\vec{\tau})s_0  +\tau_z(\vec{\alpha}_{\vec{k}}\cdot \vec{s})+ \tau_0(\vec{B}\cdot {\vec{s}}),\label{eq:ham}
\end{align}
where the $\tau_\mu$ matrices act on the U sublattices $A$ and $B$, while $s_\mu$ denotes the spin Pauli matrices. The last term is the Zeeman coupling. Since inversion exchanges the sublattices, i.e. ${\cal P}=\tau_x$, the two intersublattice hoppings $t_{x,\vec{k}}$ and $t_{y,\vec{k}}$ are even and odd in the wavevector, respectively. The local breaking of inversion symmetry at each sublattice site permits a staggered spin-orbit coupling (SOC)  $\vec{\alpha}_{\vec{k}}$ which is also odd in $\vec{k}$. 
Following Ref.~\cite{Tatsuya2021}, we consider the following tight-binding forms
\begin{align}
\varepsilon_{0,\vec{k}}=& t_1 \cos k_x+t_2 \cos  k_y -\mu, \label{eq:tb1}\\
t_{x,\vec{k}}=& m_0 + t_{3}\cos(k_x/2)\cos(k_y/2)\cos(k_z/2),\\
t_{y,\vec{k}}=& t_{4}\cos(k_x/2)\cos(k_y/2)\sin(k_z/2),\\
\alpha_{x,\vec{k}}=& \alpha_1 \sin k_y,\quad \alpha_{y,\vec{k}}= \alpha_2 \sin k_x,\\
\alpha_{z,\vec{k}}=& \alpha_3 \sin(k_x/2)\sin(k_y/2)\sin(k_z/2). \label{eq:tb5}
\end{align}
Fitting the effective model Eq.~\eqref{eq:ham} to realistic Fermi surfaces of UTe$_2$ typically requires that $t_{x,\vec{k}}$ be the largest sublattice-nontrivial term~\cite{Tatsuya2021,Yu2023,Fujimoto2024,Ando2025}.

In previous investigations based on this Hamiltonian, the SOC terms are taken to be small or vanishing, but the physical justification for this simplification is unclear. To address this we consider a coherent heavy fermion state in UTe$_2$ that is formed from the hybridization of the U~$5f$ orbitals with the conduction electrons. We suppose the dominance of the lowest-lying doublets from the two uranium ions of the primitive unit cell in the composition of the itinerant state. To obtain a minimal effective model for the itinerant electrons associated with those doublets, we integrate out the conduction electrons from the U $6d$ or Te $5p$ orbitals, see the {\emph {End Matter}} for details. This yields a Hamiltonian of the form given above, and hence sheds light on the relative sizes of the various tight-binding parameters. 
Surprisingly, we find that $f$-$d$ hybridization generates all tight-binding parameters in Eqs~\eqref{eq:tb1}-\eqref{eq:tb5} except $m_0$, implying that the SOC terms are generically comparable to other sublattice-nontrivial terms; in contrast, the only sublattice-nontrivial terms generated by $f$-$p$ hybridization are the in-plane SOC and intradimer $m_0$ hopping. These results suggest that the SOC should be comparable to the spin-independent inter-sublattice hopping. 

Although the pairing interaction in UTe$_2$ remains controversial, here we follow Ref.~\citep{Tatsuya2021} and assume a spin-fluctuation mechanism. In particular, DFT+$U$~\cite{Tatsuya2021} and DFT-DMFT~\cite{Xu2019} calculations suggest that the 
we consider a phenomenological ferromagnetic interaction between the U atoms on each dimer. 
\begin{equation}\label{eq:Hint}
    H_{\text{int}} = -\sum_{j}\sum_{\mu=x,y,z}J_\mu S^\mu_{1,j}S^{\mu}_{2,j}
\end{equation}
where $S^{\mu}_{1(2),j}$ is the $\mu$-spin operator on the dimer site 1 (2) in unit cell $j$. We include an orthorhombic anisotropy in the ferromagnetic exchange interactions $J^\mu>0$, reflecting the crystal structure. We note that these dimer ferromagnetic interactions are consistent with the low energy magnetic excitations observed by neutron scattering \cite{Knafo:2025}. Decoupling the interaction Eq.~\eqref{eq:Hint} in the Cooper channel favours intra-dimer spin-triplet pairing states with matrix pairing potential $\Delta \tau_y s_\nu$. The bare coupling constant in the $\tau_ys_\nu$ channel is $\lambda_\nu = -J_\nu + \sum_{\mu\neq \nu} {J}_\mu$. Based on the hierarchy of normal-state susceptibilities indicating the easy-axis along the $a$-axis~\citep{Miyake2019,Ran2019a}, we expect that $J_x\gg J_y\approx J_z$, which favours the $\tau_y s_y$ and $\tau_y s_z$ pairing channels.  
Despite having no momentum-dependence, the $\tau_ys_\nu$ states are odd under inversion, with both the parity and fermionic antisymmetry encoded in the $\tau_y$ sublattice-dependence. The $\tau_ys_\nu$ states belong to the $B_{2u}$, $B_{3u}$, and $A_u$ irreducible representations (irreps) of the $D_{2h}$ point group for $\nu = x$, $y$, and $z$, respectively.

The critical temperature of the $\tau_ys_\nu$ pairing states are given by the solution of $1=\lambda_\nu \chi_{\nu}$, where the pairing susceptibility is defined $\chi_\nu = \sum_{m,m'}[\chi_\nu]^{m}_{m'}$, with
\begin{equation}[\chi_\nu]^{m}_{m'} = \frac{1}{\beta N}\sum_{\vec{k}}\sum_{i\omega_n}\frac{[\tau_ys_\nu]^m_{m'}}{(i\omega_n - \epsilon_{\vec{k},m})(i\omega_n + \epsilon_{-\vec{k},m'})}
\end{equation}
Here we introduce the matrix elements between the different eigenstates $|\vec{k},m\rangle$ of the Hamiltonian
\begin{equation}
[\tau_ys_\nu]^m_{m'} = |\langle \vec{k},m| \tau_y s_\nu (is_y){\cal K}|
-\vec{k},m'\rangle|^2 \label{eq:matrixel}
\end{equation}
where ${\cal K}$ is complex conjugation. Due to the sublattice-spin texture of the normal-state bands, generally both intraband ($\epsilon_{\vec{k},m} = \epsilon_{\vec{k},m'}$) and interband ($\epsilon_{\vec{k},m} \neq  \epsilon_{\vec{k},m'}$) pairing matrix elements are nonzero. The former is  proportional to the gap at the Fermi surface and thus ensures the existence of a weak-coupling instability. The susceptibility also includes a weakly-temperature-dependent contribution from pairing involving states away from the Fermi surface, which arises due to the absence of a cutoff in our model interaction.

In the absence of a magnetic field, the sum of the matrix elements over the Kramers-degenerate states evaluates to $2[1 - (t_{x,\vec{k}}^2 + \alpha_{\nu,\vec{k}}^2)/(|\vec{t}_{\vec{k}}|^2 + |\vec{\alpha}_{\vec{k}}|^2)] \leq 2$. This so-called superconducting fitness~\cite{Ramires2018} takes its maximum (minimum) value when there is only pairing between quasiparticles in degenerate (nondegenerate) states  at $\vec{k}$ and $-\vec{k}$. Since the Cooper log is multiplied by the Fermi-surface average of this quantity, the critical temperature sensitively depends upon its value. We observe that the $t_{x,\vec{k}}$ hopping and $\alpha_{\nu,\vec{k}}$ spin-orbit coupling reduce the fitness; since $t_{x,\vec{k}}$ is likely the largest term,  the critical temperature of the $\tau_y s_\nu$ states is much smaller than naively expected from the interaction constant.

\section{Reentrant superconductivity}

\begin{figure}
\includegraphics[width=1.0\linewidth]{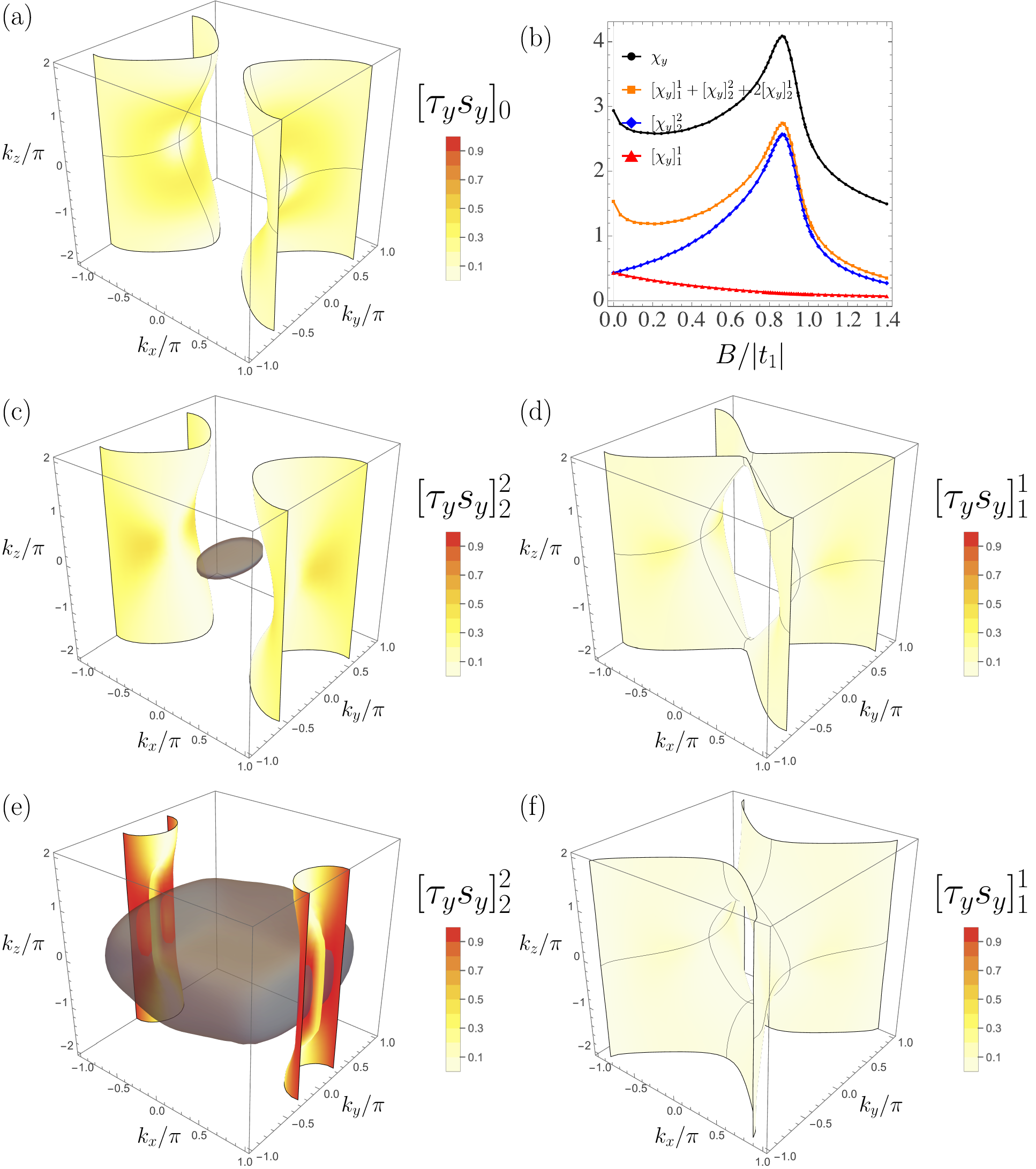}
\caption{\label{fig:fig2}Hole pockets of UTe$_2$ obtained by using the hopping amplitudes
$(t_1,t_2)=(-1, 0.76)$, $(m_0,t_3,t_4)=(-0.87, 0.83, -0.83)$, and $(\alpha_1,\alpha_2,\alpha_3)=(0.448, 0.224, 0.112)$ at (a) zero field, (b-c) a finite Zeeman field $B_y=0.2|t_1|$, and (d-e) $B_y=0.75|t_1|$. $\mu=1.625$ is used for (a) and $\mu$ for (b) is determined by the number conservation. The color gradient in (a) depicts the half of the total pairing susceptibility $[\tau_y s_y]_0$ defined at zero-field. The gray surfaces in (c) and (e) represent the anticrossing surface: pairing is significantly enhanced when this intersects the Fermi surface, as shown in (e).}
\end{figure}

Switching on the Zeeman field initially suppress the $\tau_y s_\nu$ states via a Pauli limiting mechanism, as they generically involve pairing between states which are spin-split by the field~\cite{Cavanagh2023}. This is evident in Fig.~\ref{fig:Fig1} and Fig.~\ref{fig:fig2}(b) as the sharp drop-off in the pairing susceptibility with increasing field for the $\tau_ys_y$ state. The anisotropic response, with markedly less suppression for fields along the $a$-axis, is consistent with a psuedospin description of the pairing at the Fermi surface, which gives a ${\bf d}$-vector lying mainly in the $y$-$z$ plane. 

At much larger field strengths, however, we observe a pronounced \emph{maximum} in the pairing susceptibility for fields near to the $y$-axis. As shown in Fig.~\ref{fig:fig2}(b), this is driven entirely by the pairing on one of the spin-split bands. At this maximum, the susceptiblity can significantly exceed the zero-field value, thus driving reentrant superconductivity. Qualitatively similar results are robustly obtained in different parameter sets, and also for the $\nu=x,z$ pairing state when the field is directed along the $\nu$-axis  (See the Supplemental Material~\citep{SM}).

To understand this result, we consider the evolution of the eigenstates of $h(\vec{k})$ with magnetic field strength. Although general solutions for the eigenstates can be found (see {\emph{End Matter}}), greater insight is obtained by sequentially switching on the nontrivial terms in Eq.~\eqref{eq:ham}. Motivated by the SC2 phase in UTe$_2$, we set $\vec{B}$ along the $y$-direction. We consequently take the $y$-axis as the quantization axis in spin space, which implies that the $\tau_ys_y$ state describes ``opposite-spin'' pairing, whereas the $\tau_ys_{\nu=x,z}$ states are ``equal-spin'' pairing.

\begin{figure}
\includegraphics[width=1.0\linewidth]{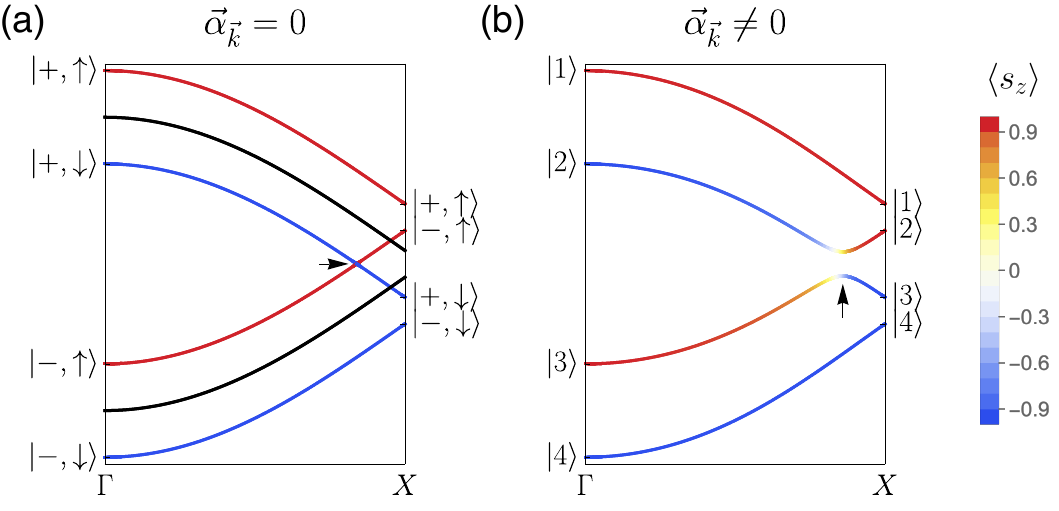}
\caption{\label{fig:fig3}A schematic of the band structure from the Hamiltonian $h_{\vec{k}}$ with the Zeeman coupling along the $\Gamma$ point and a time-reversal invariant momentum $X$ in the first Brillouin zone. The black lines depicts the two two-fold degenerate bands at zero Zeeman field while the color-coded lines represent the electronic bands at a finite Zeeman field. Red and blue indicate that the average spin polarization of the corresponding electronic state is aligned parallel and antiparallel to the Zeeman field, respectively, while the intermediate values are represented by a color gradient. The arrows mark the crossing (anticrossing) of two middle bands originating from $|+,\downarrow\rangle$ and $|B,\uparrow\rangle$ at the $\Gamma$ point in the absence (presence) of spin-orbit coupling.}
\end{figure}

We start by assuming that only $t_{x,\vec{k}}$ is nonzero, which is a reasonable starting point as we expect this to be the largest nontrivial term.  In the absence of the Zeeman field, the two spin-degenerate bands correspond to the bonding and antibonding states on the U dimer, i.e. $|+,\sigma\rangle = \frac{1}{\sqrt{2}}(|A\rangle + |B\rangle)\otimes|\sigma\rangle$ and $|-,\sigma\rangle = \frac{1}{\sqrt{2}}(|A\rangle - |B\rangle)\otimes|\sigma\rangle$, respectively. Consistent with the vanishing superconducting fitness, the $\tau_ys_\nu$ states are all interband, as the odd-parity sublattice-dependence enforces pairing between electrons in bonding and antibonding states. The Zeeman field lifts the spin-degeneracy, with spin-$\uparrow$ and $\downarrow$ states indicated by the red and blue lines in Fig.~\ref{fig:fig3}(a), respectively.  Notably, the states $|-,\uparrow\rangle$ and $|+,\downarrow\rangle$ are shifted  toward one another, and intersect on the surface $|t_{x,\vec{k}}|=B$ when the Zeeman field exceeds the minimum band splitting at zero field. The existence of this crossing is essential to our theory. 

We now turn on the spin-flip SOC $\vec{\alpha}_{\perp,\vec{k}}=(\alpha_{x,\vec{k}},0,\alpha_{z,\vec{k}})$, which hybridizes the $|\pm,\sigma\rangle$ states; due to the absence of a simple quantum number, we henceforth label them as $n=1\,\ldots,4$ in descending order of energy. The highest-energy $n=1$ state is 
\begin{equation}
|1\rangle = \cos(\tfrac{1}{2}\zeta_+)|+,\uparrow\rangle + e^{i\phi_\alpha}\sin(\tfrac{1}{2}\zeta_+)|-,\downarrow\rangle
\end{equation}
where $\phi_\alpha$ is the angle that $\vec{\alpha}_{\perp,\vec{k}}$ makes with the $x$-axis, and $\tan\zeta_+ = |\vec{\alpha}_{\perp,\vec{k}}|/(|t_{x,\vec{k}}|+B)$. The SOC mixes the outermost states of the $t_{x,\vec{k}}$-only system, but 
for weak SOC the highest-energy state maintains a predominantly $|+,\uparrow\rangle$-character, which is enhanced by increasing field. Importantly, the SOC also generates a nonzero intraband matix element $[\tau_ys_y]^{1}_{1}=\sin^2\zeta_{+}$, and thus allows a weak-coupling instablity which is nevertheless suppressed by the field. On the other hand, the intraband matrix elements of the equal-spin $\tau_{y}s_{\nu=x,z}$ states remain zero, as it is still not possible to pair bonding and antibonding states with the same spin. A similar analysis holds for the lowest-energy $|4\rangle$ state.

The hybridization of the $|+,\downarrow\rangle$ and $|-,\uparrow\rangle$ states by the spin-flip SOC turns their crossing into an anticrossing, see Fig.~\ref{fig:fig3}(b). This is reflected in the wavefunction of the second state 
\begin{equation}
|2\rangle = \cos(\tfrac{1}{2}\zeta_-)|+,\downarrow\rangle - e^{i\phi_\alpha}\sin(\tfrac{1}{2}\zeta_-)|-,\uparrow\rangle
\end{equation}
The angle $\zeta_-$, defined $\tan\zeta_-=|\vec{\alpha}_{\perp,\vec{k}}|/(|t_{x,\vec{k}}|-B)$, sweeps from $0$ to $\pi$ as we move across the anticrossing, reflecting the change in character from $|+,\downarrow\rangle$-like to $|-,\uparrow\rangle$-like; exactly at the anticrossing we have $\zeta_-=\frac{\pi}{2}$, implying an equal superposition of the original states. The matrix element of the opposite-spin pairing state is  $[\tau_ys_y]^{2}_{2}=\sin^2(\zeta_-)$, which reaches the maximal possible value of $1$ exactly at the anticrossing. 
Although the matrix element evaluates to one only on the anticrossing surface $|t_{x,\vec{k}}|=B$, its value remains elevated within a  momentum shell of width $\delta k\sim |\vec{\alpha}_{\vec{k}}|/|\vec{v}^0_{\vec{k}}|$ around the anticrossing, where $\vec{v}^{0}_{\vec{k}}$ is the velocity of the unhybridized bands at $\vec{k}$. If the anticrossing surface intersects the Fermi surface of bands 2 or 3, we therefore expect a large enhancement of the pairing susceptibility near the intersection line. 
Again, the equal-spin pairing states have vanishing intraband matrix elements.

Including the remaining terms $t_{y,\vec{k}}$ and $\alpha_{y,\vec{k}}$ complicates the analysis, as now all basis states $|\pm,\sigma\rangle$ are represented in each eigenstate. However, the physics outlined above remains mainly intact. Specifically, the anticrossing persists but is shifted to $B =\sqrt{t_{x,\vec{k}}^2+t_{y,\vec{k}}^2+\alpha_{y,\vec{k}}^2}$, while the matrix element for opposite-spin pairing on the middle two bands takes a maximum on this surface but with reduced value $[\tau_ys_z]^{2}_{2}={(t_{x,\vec{k}}^2 + \alpha_{y,\vec{k}}^2)}/{(t_{x,\vec{k}}^2 + t_{y,\vec{k}}^2+\alpha_{y,\vec{k}}^2)}$. So long as $t_{y,\vec{k}}$ is small compared to the other terms, however, the matrix element will remain close to its maximal value. Meanwhile, the equal-spin pairing states are able to open gaps on the Fermi surface, but display only weak magnetic-field dependence.

We can now understand the origin of the maximum in the pairing susceptibility in Fig.~\ref{fig:Fig1}. 
For the chosen tight-binding parameters only a single band crosses the Fermi surface at zero field. Since $t_{x,\vec{k}}$ is the dominant term in the Hamiltonian, the superconducting fitness of the $\tau_ys_y$ state is small over the Fermi surface, as shown in Fig.~\ref{fig:fig2}(a). Switching on a Zeeman field along the $y$-axis, we first suppress the pairing susceptibility as we lose the Cooper log contributed by pairing between the spin-split bands. However, the field also modifies the intraband matrix elements for the spin-split bands, reducing them on one of the bands ($n=1$) but enhancing them on the other ($n=2$); as can be seen in Fig.~\ref{fig:fig2}(b), this effect tends to cancel out at small fields, but if the matrix elements at zero field are sufficiently small, the enhancement on band 2 can overcompensate the reduction on band 1. This enhancement already occurs without the intersection of the avoided crossing surface with band 2, see Fig.~\ref{fig:fig2}(c) and (d). The matrix elements on band 2 take their maximum value at the intersection with the anticrossing surface (Fig.~\ref{fig:fig2}(e)), but remain elevated over the entire Fermi surface due to the effect of the spin-flip SOC, thus giving rise to the broad peak in the pairing susceptibility; at this point the gap is heavily suppressed on band 1, as shown in Fig.~\ref{fig:fig2}(f). The reduction at higher fields occurs after the anticrossing has passed over the entire Fermi surface, and the increasing Zeeman splitting pushes band 2 below the Fermi level. 
For fields perpendicular to the $y$-axis, the behaviour is essentially the same as for the $\nu=x$, $z$ states with $y$-axis field, and so there is no enhancement of the intraband matrix elements and the pairing suscepibility monotonically decreases with field strength.

\section{Discussion}

The peak in the pairing susceptibility  for the $B_{3u}$-symmetry 
$\tau_ys_y$ pairing state shown in Fig.~\ref{fig:Fig1} recalls the location of the SC2 phase in UTe$_2$ at ambient pressure. Although our theory is based on a highly-simplified description of the $f$-electrons, to be a plausible explanation for the SC2 phase it is necessary that there is at least approximate consistency between the energy scales of our model and experiment. 

The maximum critical temperature of the SC2 state occurs at a field strength of about $30$~T. Equating this with the peak in the susceptibility implies that the hopping $|t_1|\approx \mu_0B = 2$~meV; the bandwidth $W$ is then roughly an order of magnitude larger, $W\sim 10$~meV. 
This estimate for the bandwidth is about $10$-$100$ times smaller than predicted by DFT+U calculations~\cite{Xu2019,Ishizuka:2019,Tatsuya2021}. Although this method gives good agreement with the Fermi surface observed in magnetic quantum oscillation measurements~\citep{Aoki2023dHvA,Eaton2024} and angle-resolved photoelectron spectroscopy~\citep{Miao2020}, it does not account for the Kondo effect. Methods which include this, such as DFT+DMFT~\citep{Xu2019,Choi2024,Sundermann2025}, suggest a much flatter quasiparticle band. A crude estimate for its bandwidth can be obtained by equating it with the Kondo temperature $T_K$: defining this by the maximum in the susceptibility along the $b$-axis, we have $T_K\approx 35$~K (i.e. $W\sim 3$~meV)~\citep{Ran2019a,Jiao2020,Eo2022,Azari2025}, which is in better agreement with the bandwidth estimate from our model.

The identification of SC2 as belonging to the $B_{3u}$ irrep also constrains the symmetry of the low-field SC1 phase. Specifically, the observation of a  phase boundary between SC1 and SC2 for magnetic fields applied along the $b$-axis~\citep{Valiska2026,Saka:2023,Kinjo2023} requires that the two phases possess different symmetries. This immediately rules out the $B_{1u}$ irrep, as it mixes with $B_{3u}$ in a $b$-axis field, yielding a crossover instead of a phase transition. The remaining possibilities are therefore $A_u$ and $B_{2u}$; the former is consistent with thermal transport and nuclear magnetic resonance~\citep{Matsumura2023,Suetsugu2024}, while the latter is suggested by specific heat and ultrasound measurements~\citep{Theuss2024,SLee2025}. 
Additionally, experiments show that pressure causes the SC2 phase to replace SC1 at zero field, while also suppressing the reentrance~\citep{Vasina2025,Wu2025,Kamat2026}. This behavior could be accounted for within our theory if pressure increases the inter-rung hopping $t_y$ in Eq.~\eqref{eq:ham}, as this term both enhances the zero-field critical temperature of the $\tau_ys_y$ state and reduces the matrix elements at the avoided crossing.

Finally, we wish to comment on the wider applicability of our results. Prior to the discovery of UTe$_2$, the most famous examples of reentrant superconductivity were the ferromagnetic superconductors URhGe and UCoGe~\citep{Aoki2019ReviewFMSC}. Although the relationship of these materials to UTe$_2$ remains controversial, the key ingredients of our theory may be present: ferromagnetic fluctuations, non-negligible spin-orbit coupling, and the dominance of the even-parity part of intersublattice hopping over the odd-parity part. Indeed, we find that the latter two conditions generically hold in $\vec{k}\cdot\vec{p}$ theories~\citep{Suh2023} about certain high-symmetry points in the Brillouin zones of UCoGe and URhGe.

\section{Conclusions}

We have proposed a theoretical explanation for the high-field SC2 phase in UTe$_2$. Specifically, we find that reentrant superconductivity in strong Zeeman fields can counterintuitively arise from opposite-spin Cooper pairing of electrons localized on the U dimers in each unit cell. Crucial to our argument is non-negligible spin-orbit coupling, which can arise from the hybridization of the localized $f$ levels with the conduction electron bands.
Our analysis reveals that the mechanism for the reentrant superconductivity is the interplay of the coupled spin-sublattice degrees of freedom in the Cooper pairs with the reconstruction of the band wavefunctions due to the Zeeman splitting. In the presence of spin-orbit coupling, this can dramatically boost the gap magnitude on one of the spin-split Fermi surfaces, thus giving a maximum in the critical temperature as a function of the Zeeman field strength. Our theory naturally explains the strong field-angle anisotropy of the phase diagram of UTe$_2$, and is also consistent with the relevant energy scales. 

{\textit Acknowledgements---} PMRB thanks E. Hassinger for stimulating discussions. CL, NAH, and PMRB were supported by the
Marsden Fund Council from Government funding, managed by Royal Society Te Ap\={a}rangi, Contract No. UOO2222. DFA was supported
by the Simons Foundation under Grant No. SFI-MPS
NFS-00006741-02 and by the Department of Energy, Office
of Basic Energy Science, Division of Materials Sciences
and Engineering under Award No. DE-SC0021971.

\bibliography{ref}

\clearpage\pagebreak{}

\appendix

\section*{End Matter}

\subsection{Eigenvectors of the Hamiltonian in Eq.~\eqref{eq:ham}}

The eigenenergies of the Hamiltonian in Eq.~\eqref{eq:ham} with the Zeeman field $B$ along the $y$-axis are given by 
\begin{align}
\xi_{1,\vec{k}}=&\varepsilon_{0,\vec{k}}+\sqrt{(\varepsilon_{\parallel,\vec{k}}+B)^2+\alpha^2_{\perp,\vec{k}}},\\
\xi_{2,\vec{k}}=&\varepsilon_{0,\vec{k}}+\sqrt{(\varepsilon_{\parallel,\vec{k}}-B)^2+\alpha^2_{\perp,\vec{k}}},\\
\xi_{3,\vec{k}}=&\varepsilon_{0,\vec{k}}-\sqrt{(\varepsilon_{\parallel,\vec{k}}-B)^2+\alpha^2_{\perp,\vec{k}}},\\
\xi_{4,\vec{k}}=&\varepsilon_{0,\vec{k}}-\sqrt{(\varepsilon_{\parallel,\vec{k}}+B)^2+\alpha^2_{\perp,\vec{k}}},
\end{align}
with $\varepsilon_{\parallel,\vec{k}}=\sqrt{t_{x,\vec{k}}^2+t_{y,\vec{k}}^2+\alpha_{y,\vec{k}}^2}$, $\alpha_{\perp,\vec{k}}=\sqrt{\alpha_{x,\vec{k}}^2+\alpha_{z,\vec{k}}^2}$. The corresponding eigenvectors are written as
\begin{align}
|1\rangle= & \cos\frac{\zeta_{+}}{2}\cos\frac{\xi}{2}|+,\uparrow\rangle+e^{i\phi_{\alpha}}\sin\frac{\zeta_{+}}{2}\cos\frac{\xi}{2}|-,\downarrow\rangle\\
 & +\sin\frac{\zeta_{+}}{2}\sin\frac{\xi}{2}|+,\downarrow\rangle+e^{i\phi_{\alpha}}\sin\frac{\zeta_{+}}{2}\sin\frac{\xi}{2}|-,\uparrow\rangle,\nonumber \\
|2\rangle= & -e^{i\phi_{\alpha}}\sin\frac{\zeta_{-}}{2}\sin\frac{\xi}{2}|+,\uparrow\rangle-\cos\frac{\zeta_{-}}{2}\sin\frac{\xi}{2}|-,\downarrow\rangle\\
 & +\cos\frac{\zeta_{-}}{2}\cos\frac{\xi}{2}|+,\downarrow\rangle+e^{-i\phi_{\alpha}}\sin\frac{\zeta_{-}}{2}\cos\frac{\xi}{2}|-,\uparrow\rangle,\nonumber \\
|3\rangle= & e^{-i\phi_{\alpha}}\cos\frac{\zeta_{-}}{2}\sin\frac{\xi}{2}|+,\uparrow\rangle-\sin\frac{\zeta_{-}}{2}\sin\frac{\xi}{2}|-,\downarrow\rangle\\
 & +\sin\frac{\zeta_{-}}{2}\cos\frac{\xi}{2}|+,\downarrow\rangle-e^{-i\phi_{\alpha}}\cos\frac{\zeta_{-}}{2}\cos\frac{\xi}{2}|-,\uparrow\rangle,\nonumber \\
|4\rangle= & \sin\frac{\zeta_{+}}{2}\cos\frac{\xi}{2}|+,\uparrow\rangle-e^{i\phi_{\alpha}}\cos\frac{\zeta_{+}}{2}\cos\frac{\xi}{2}|-,\downarrow\rangle\\
 & -e^{i\phi_{\alpha}}\cos\frac{\zeta_{+}}{2}\sin\frac{\xi}{2}|+,\downarrow\rangle+\sin\frac{\zeta_{+}}{2}\sin\frac{\xi}{2}|-,\uparrow\rangle,\nonumber 
\end{align}
with $|\pm,\sigma\rangle=\frac{1}{\sqrt{2}}\{e^{-\frac{i\phi_{t}}{2}}|A,\sigma\rangle\pm e^{\frac{i\phi_{t}}{2}}|B,\sigma\rangle\}$. The momentum-dependence suppressed for conciseness. Here, $\tan\zeta_\pm\equiv\alpha_\perp/(\varepsilon_\parallel\pm B)$, and 
\begin{align}
\varepsilon_{\parallel,\vec{k}}(\cos \xi_{\vec{k}},\sin \xi_{\vec{k}})=&(\sqrt{t_{x,\vec{k}}^2+t_{y,\vec{k}}^2},\alpha_{y,\vec{k}}),\\
(t_{x,\vec{k}},t_{y,\vec{k}})=&\sqrt{t_{x,\vec{k}}^2+t_{y,\vec{k}}^2}(\cos \phi_{t,\vec{k}},\sin \phi_{t,\vec{k}}),\\
(\alpha_{x,\vec{k}},\alpha_{z,\vec{k}})=&\alpha_{\perp,\vec{k}}(\cos \phi_{\alpha,\vec{k}},\sin \phi_{\alpha,\vec{k}}).
\end{align}

Note that $\phi_{t,\vec{k}},\,\xi_{\vec{k}},\rightarrow0$ in the
limit of $t_{y,\vec{k}}\rightarrow0$ and $\alpha_{y,\vec{k}}\rightarrow0$,
which results in 
\begin{align}
|1\rangle= & \cos\frac{\zeta_{+}}{2}|+,\uparrow\rangle+e^{i\phi_{\alpha}}\sin\frac{\zeta_{+}}{2}|-,\downarrow\rangle,\\
|2\rangle= & \cos\frac{\zeta_{-}}{2}|+,\downarrow\rangle+e^{-i\phi_{\alpha}}\sin\frac{\zeta_{-}}{2}|-,\uparrow\rangle,\\
|3\rangle= & \sin\frac{\zeta_{-}}{2}|+,\downarrow\rangle-e^{-i\phi_{\alpha}}\cos\frac{\zeta_{-}}{2}|-,\uparrow\rangle,\\
|4\rangle= & \sin\frac{\zeta_{+}}{2}|+,\uparrow\rangle-e^{i\phi_{\alpha}}\cos\frac{\zeta_{+}}{2}|-,\downarrow\rangle.
\end{align}

\subsection{Spin-orbit coupling generated by hybridization with the conduction bands}
\begin{table*}[t]
\centering
\footnotesize
\caption{Minimal-model parameters in terms of the $f$-$d$ and $f$-$p$ hybridisation parameters, obtained by matching the momentum form factor in each $\tau_i\sigma_j$ channel. For the $p_y$ orbitals we expand in $1/\Delta_{pf}$ with $\Delta_{pf}\equiv\epsilon_p-\epsilon_f$ and keep the leading order contribution.}
\label{tab:minimal_model}
\setlength{\tabcolsep}{3pt}
\renewcommand{\arraystretch}{1.2}
\resizebox{\textwidth}{!}{
\begin{tabular}{@{}c c l l@{}}
\toprule
parameter & $\tau_i\sigma_j$, form factor & $V^{fd}$ contribution & $V^{fp}$ contribution \\
\midrule
$-\mu$ & $\tau_0\sigma_0$, $1$ & $\begin{aligned}[t]
\tfrac{1}{\Delta_{fd}}\bigl[&4(t_{d,0}^{fd})^2+4(t_{d,x}^{fd})^2+4(t_{d,y}^{fd})^2+4(t_{d,z}^{fd})^2\\
&+2(t_{x,y}^{fd})^2+2(t_{y,x}^{fd})^2+(t_{z,0}^{fd})^2\bigr]
\end{aligned}$ & $-\tfrac{4}{\Delta_{pf}}\bigl[(t_0^{fp})^2+(t_x^{fp})^2+(t_y^{fp})^2+(t_z^{fp})^2\bigr]$ \\
\midrule
$t_1$ & $\tau_0\sigma_0$, $\cos k_x$ & $\tfrac{4}{\Delta_{fd}}\bigl[(t_{d,0}^{fd})^2+(t_{d,x}^{fd})^2-(t_{d,y}^{fd})^2-(t_{d,z}^{fd})^2\bigr]$ & $-\tfrac{4}{\Delta_{pf}}\bigl[(t_0^{fp})^2+(t_x^{fp})^2-(t_y^{fp})^2-(t_z^{fp})^2\bigr]$ \\
\midrule
$t_2$ & $\tau_0\sigma_0$, $\cos k_y$ & $\tfrac{4}{\Delta_{fd}}\bigl[(t_{d,0}^{fd})^2-(t_{d,x}^{fd})^2+(t_{d,y}^{fd})^2-(t_{d,z}^{fd})^2\bigr]$ & $0$ \\
\midrule
$m_0$ & $\tau_x\sigma_0$, $1$ & $0$ & $\tfrac{4}{\Delta_{pf}}\bigl[(t_0^{fp})^2-(t_x^{fp})^2-(t_y^{fp})^2+(t_z^{fp})^2\bigr]$ \\
\midrule
$t_3$ & $\tau_x\sigma_0$, $\cos\tfrac{k_x}{2}\cos\tfrac{k_y}{2}\cos\tfrac{k_z}{2}$ & $\tfrac{8}{\Delta_{fd}}\bigl[t_{d,x}^{fd}\,t_{y,x}^{fd}+t_{d,y}^{fd}\,t_{x,y}^{fd}\bigr]$ & $0$ \\
\midrule
$t_4$ & $\tau_y\sigma_0$, $\cos\tfrac{k_x}{2}\cos\tfrac{k_y}{2}\sin\tfrac{k_z}{2}$ & $\tfrac{8}{\Delta_{fd}}\bigl[t_{d,x}^{fd}\,t_{y,x}^{fd}+t_{d,y}^{fd}\,t_{x,y}^{fd}\bigr]$ & $0$ \\
\midrule
$\alpha_1$ & $\tau_z\sigma_x$, $\sin k_y$ & $\tfrac{8}{\Delta_{fd}}\bigl[t_{d,0}^{fd}\,t_{d,x}^{fd}-t_{d,y}^{fd}\,t_{d,z}^{fd}\bigr]$ & $\tfrac{8\,t_{y,2}^p}{\Delta_{pf}^{\,2}}\bigl[t_0^{fp}\,t_x^{fp}-t_y^{fp}\,t_z^{fp}\bigr]$ \\
\midrule
$\alpha_2$ & $\tau_z\sigma_y$, $\sin k_x$ & $\tfrac{8}{\Delta_{fd}}\bigl[t_{d,0}^{fd}\,t_{d,y}^{fd}+t_{d,x}^{fd}\,t_{d,z}^{fd}\bigr]$ & $\tfrac{8}{\Delta_{pf}}\bigl[t_0^{fp}\,t_y^{fp}+t_x^{fp}\,t_z^{fp}\bigr]$ \\
\midrule
$\alpha_3$ & $\tau_z\sigma_z$, $\sin\tfrac{k_x}{2}\sin\tfrac{k_y}{2}\sin\tfrac{k_z}{2}$ & $\tfrac{8}{\Delta_{fd}}\,t_{d,z}^{fd}\,t_{z,0}^{fd}$ & $0$ \\
\bottomrule
\end{tabular}
}
\end{table*}

Here, we show that the hybridization of localized $f$-electrons with the conduction electrons is sufficient to generate significant spin-orbit coupling, and as such, cannot be neglected in minimal models. As an illustrative example, we consider a model for ${\rm UTe}_2$ with localized $J=5/2$ $f$-electron states on the ${\rm U}$ atoms (Wyckoff position $4i$ of the space group No. 71) and $p_y$ and $d_{xy}$ conduction electrons on the ${\rm Te}(2)$ (Wyckoff position $4h$) and ${\rm U}$ atoms, respectively. We assume the $f$-electron physics near the Fermi surface is dominated by a single Kramers doublet~\cite{Tatsuya2021}. In the presence of an orthorombic crystal electric field, this takes the form~\cite{Theuss2024}
\begin{align}
|\Gamma^\pm\rangle = \alpha |\pm 5/2\rangle + \beta |\pm 1/2\rangle + \gamma |\mp 3/2\rangle ,
\end{align}
with $|\alpha|^2+|\beta|^2+|\gamma|^2=1$.
We determine the non-local hybridization matrix using the Slater-Koster two-centered integral approach, whereby the spin-orbit coupled $f$-electron ground state naturally introduces non-zero overlaps in both the spin-preserving and spin-flipping hybridization.

The $f$-$d$ hybridization is given by
\begin{align}
\hat{V}=\sum_{\vec{k}}\sum_{l_{ f},s_f}\sum_{l_d,s_d}[V^{fd}_{\vec{k}}]_{l_{\rm f} s_f;l_{d} s_d}\hat{f}^{\dagger}_{l_{f},s_f}\hat{d}_{l_{ d},s_d},
\end{align}
where the summation is carried over the two sublattices $l_f, l_d={A,B}$ of U atoms and the two Kramers degrees of freedom $s_f=\pm$ ($s_d=\uparrow,\downarrow$). $\hat{f}_{l_{f},s_f}$ and $\hat{d}_{l_{ d},s_d}$ are the annihilation operators for the $f$ and $d$ electrons. The hybridization matrix $V^{fd}_{\vec{k}}$ is written as
\begin{equation}
  V^{fd}_{\vec{k}} = \begin{pmatrix}
    V_{\text{in},\vec{k}}  & V_{BA,\vec{k}} + t_{z,0}^{fd}\sigma_0 \\
    V_{AB,\vec{k}}  - t_{z,0}^{fd}\sigma_0  & V_{\text{in},\vec{k}}
  \end{pmatrix},
\end{equation}
with
\begin{align}
V_{\text{in},\vec{k}} =& -2t_{x,y}^{fd}\sin k_x\sigma_y - 2t_{y,x}^{fd}\sin k_y\sigma_x, \\
V_{AB,\vec{k}}  = &\,4e^{ik_z/2}\Big[
\cos \frac{k_y}{2} (t_{d,0}^{fd}\cos \frac{k_x}{2}\sigma_0 - t_{d,y}^{fd} \sin \frac{k_x}{2}\sigma_y) \nonumber\\
&- \sin \frac{k_y}{2}(t_{d,x}^{fd}\cos \frac{k_x}{2} \sigma_x + it_{d,z}^{fd}\sin \frac{k_x}{2} \sigma_z)\Big], \\
V_{BA,\vec{k}} = &-4e^{-ik_z/2}\Big[
\cos \frac{k_y}{2}(t_{d,0}^{fd}\cos \frac{k_x}{2}\sigma_0 +t_{d,y}^{fd} \sin \frac{k_x}{2}\sigma_y) \nonumber\\
&+\sin \frac{k_y}{2}(t_{d,x}^{fd}\cos \frac{k_x}{2} \sigma_x - it_{d,z}^{fd}\sin \frac{k_x}{2}\sigma_z)\Big],
\end{align}
where the elements associated with the Pauli matrices $\sigma_{0,z}$ ($\sigma_{x,y}$) represent spin-preserving (spin-flipping) hybridization. $V_{\text{in}}$ originates from in-plane intrasublattice bonds, $V_{AB}$ and $V_{BA}$ from intersublattice diagonal bonds, and $t_{z,0}^{fd}$ from the intersublattice bonds along the $z$-direction within the primitive unit cell. Each $t^{fd}$ coefficient arises from symmetry allowed non-local hybridization between the $\Gamma^{\pm}$ doublets and the $d_{xy}$-orbitals at the U atoms.

The $f$-$p$ hybridization is given by
\begin{align}
\hat{V}=\sum_{\vec{k}}\sum_{l_{ f},s_f}\sum_{l_d,s_d}[V^{fp}_{\vec{k}}]_{l_{\rm f} s_f;l_{p} s_p}\hat{f}^{\dagger}_{l_{f},s_f}\hat{p}_{l_{ p},s_p},
\end{align}
where the summation is carried over the two sublattices $l_f={A,B}$ ($l_p={A,B}$) of U atoms (Te(2) atoms) and the two Kramers degrees of freedom $s_f=\pm$ ($s_p=\uparrow,\downarrow$). $\hat{f}_{l_{f},s_f}$ and $\hat{p}_{l_{p},s_p}$ are the annihilation operators for the $f$ and $p$ electrons. The hybridization matrix $V^{fp}_{\vec{k}}$ is written as
\begin{align}
V^{fp}_{\vec{k}} = &(e^{-ik_x}-1)[t_{y}^{fp}(\tau_y{+}i\tau_z)\sigma_y + t_{z}^{fp}(\tau_y{-}i\tau_z)\sigma_z]\\
&-(e^{-ik_x}+1)[t_{0}^{fp}(\tau_0-\tau_x)\sigma_0 - it_{x}^{fp}(\tau_0+\tau_x)\sigma_x],\nonumber
\end{align}

To determine the effective low-energy theory for the $f$-electrons, we integrate out the conduction electrons. For $d(p)$-electrons, the effective Hamiltonian is given by
\begin{align}
H_\text{eff} (\vec{k}) = \epsilon_{f} + V^{fd(p)}_{\vec k} G_{d(p)}(\vec{k}, i\omega_n \rightarrow \epsilon_f)  V^{fd(p)\dagger}_{\vec k},\label{eq:H_eff}
\end{align}
where $\varepsilon_{f}$ is the energy level of the local $\Gamma^{\pm}$ states, and $G_{d(p)}(\vec{k}, \epsilon_f )$ is the Green function for the $d(p)$-electrons which we have analytically continued to the real frequency $\epsilon_{f}$. For $d$-electrons, it is sufficient to consider terms generated by the trivial part, which is proportional to the identity matrix, of the Green function $G_{d}(\vec{k}, \epsilon_{f})$ to derive the spin-orbit couplings in Eq.~\eqref{eq:ham}.

For the $p$-electrons, the Green's function is given by $G_{p}({\vec{k}}, \epsilon_{f}) = [\epsilon_{f} - H^{p}_{\vec{k}}]^{-1}$, where the Hamiltonian $H^{p}_{\vec{k}}$ for the $p$-electrons from Te(2) atoms is written as
\begin{equation}
H^{p}({\vec{k}}) = \begin{pmatrix}
\epsilon_p + 2 t_{x}^p\cos k_x & t_{y,1}^p + t_{y,2}^p\, e^{-i k_y} \\
t_{y,1}^p + t_{y,2}^p\, e^{+i k_y} & \epsilon_p + 2 t_{x}^p\cos k_x
\end{pmatrix}\otimes\sigma_0,
\end{equation}
in the nearest-neighbor hopping approximation restricted to the Te(2) sites. $\sigma_0$ represents the identity matrix in the spin space. $\epsilon_p$ is the onsite energy of the $p_y$- electrons at Te(2) sites, $t_{x}^p$ is the intrasublattice hopping along the $x$-axis, and $t_{y,1}^p$ and $t_{y,2}^p$ are the intersublattice hoppings along the $y$-axis.

By expanding the terms appearing in Eq.~\eqref{eq:H_eff} with respect to the characteristic energy scale $\Delta_{fd(p)}\equiv \epsilon_{d(f)}-\epsilon_{f}$ and comparing the results with the Hamiltonian in Eq.~\eqref{eq:ham}, we derive the expression for the hopping parameters in in Eq.~\eqref{eq:ham} in terms of the hybridization amplitudes, as summarized in Table~\ref{tab:minimal_model}. Note that the hybridization between the $f$- and the  $p$-orbitals induces just the $m_0$ term in the coefficient of the matrix $\tau_x s_0$ in Eq.~\eqref{eq:ham}.
\end{document}